\documentclass[iop]{emulateapj}

\usepackage{graphicx}

\usepackage{graphicx}

\newcommand{\hii}{H\,{\sc ii}~}

\shortauthors{Bern\'e \& Matsumoto}  
\shorttitle{The Kelvin-Helmholtz instability  in Orion}

\begin{document}

\title{The Kelvin-Helmholtz instability  in Orion: \\
 a source of turbulence and chemical mixing}

\author{%
O. Bern\'e \altaffilmark{1,2}, 
Y. Matsumoto \altaffilmark{3},
}


\altaffiltext{1}{Universit\'e de Toulouse; UPS-OMP; IRAP;  Toulouse, France}
\altaffiltext{2}{CNRS; IRAP; 9 Av. colonel Roche, BP 44346, F-31028 Toulouse cedex 4, France} 
\altaffiltext{3}{Department of Physics, Chiba University, 1-33 Yayoi-cho, Inage-ku, Chiba 263-8522, JAPAN}

\begin{abstract}

Hydrodynamical instabilities are believed to power some of the small scale (0.1-10 pc) turbulence and 
chemical mixing in the interstellar medium. Identifying such instabilities has always been difficult
but recent observations of a wavelike structure (the \emph{Ripples}) in the Orion nebula have been interpreted as a
signature of the Kelvin-Helmholtz instability (KHI), occurring at the interface between the \hii region and 
the molecular cloud. However, this has not been verified theoretically. In this letter, we investigate theoretically 
the stability of this interface using observational constraints for the local physical conditions. 
A linear analysis shows that the \hii/molecular cloud interface is indeed KH unstable for a certain 
range of magnetic field orientation. We find that the maximal growth-rates correspond to typical timescales 
of a few $10^4$ years and instability wavelengths of 0.06 to 0.6 pc. We predict that after 
$2\times10^5$ years the KHI saturates and forms a turbulent layer of about 0.5 pc. The KHI
can remain in linear phase over a maximum distance of 0.75~pc. These spatial and time scales are 
compatible with the \emph{Ripples} representing the linear phase of the KHI.
These results suggest that the KHI may be crucial to generate turbulence 
and to bring heavy elements injected by the winds of  massive stars in \hii regions to colder regions 
where planetary systems around  low mass stars are being formed. This could apply to the transport 
of $^{26}$Al injected by a massive star in an \hii region to the nascent solar-system.

\end{abstract}


\section{Introduction} \label{int}

Over 50 years ago, it was postulated by \citet{spi54} and \citet{fri54} that hydrodynamical
instabilities may form in interface regions between the hot diffuse gas ionized by massive stars
and cold dense molecular clouds.  In particular, these authors suggested that 
\emph{elephant trunk}, or spike structures, which are widely observed in star-forming 
regions, may result from the Rayleigh-Taylor (see \citealt{cha61}) instability (RTI). 
Another classical type of interface instability is the Kelvin-Helmholtz instability 
(see \citealt{cha61}) which occurs in the presence of a velocity sheer across the 
interface and is characterized by a wavelike periodic structure. Both types of instabilities 
have been considered to play a dominant role in the interstellar medium,
as a power source for small scale (0.1-10 pc) turbulence \citep{elm04} and in the 
mixing of chemical elements \citep{roy95}. Observationally, it has been hard to confirm 
the existence of these instabilities. Although the observed sizes of \emph{elephant trunks} 
match first order theoretical models of RTI \citep{fri54}, recent observations of the velocity field 
in the Pillars of Creation and the Horsehead nebula by \citet{pou98} and \citet{pou03}
tend to discard the RTI hypothesis for the formation of these structures (instead, the
selective photodissociation model of \citet{rei83} is invoked). More recently, 
\citet{ber10} (BMC hereafter) observed a periodic wavelike structure (the \emph{Ripples} hereafter), at the surface of 
the Orion cloud (Fig. 1) which appears to be compatible with a KHI. 
However, this work was mostly qualitative and lacked a detailed model to assess if 
the development of the KHI is possible in conditions as those found in Orion. In addition, this 
former study did not consider the possible effect of magnetic fields --which are known to be 
strong in Orion \citep{abe04}-- on the KHI. Finally, even if there is evidence for
the existence of the KHI in the interstellar medium (ISM), it remains unclear over which timescale this instability
may convert energy into turbulent motion of the gas and hence if it can actually play a role
in mixing the hot and cold gas. On the theoretical side,  extensive 
numerical magnetohydrodynamical (MHD) models for the KHI have been developed 
and successfully applied to explain observed phenomena in solar-system plasmas 
(e.g. \citealt{mat04}). In addition, \citet{mat10} have studied in great details the saturation 
of the KHI and its evolution into turbulence and mixing of the gas.
In this letter, we perform a linear MHD analysis applied to the situation in Orion's \emph{Ripples},
using physical parameters determined observationally. We derive the key parameters
that characterize the instability and use them to determine the timescales over which
the gas becomes turbulent due to the saturation of the instability. We discuss these results
in the context of chemical mixing in the interstellar medium and transport of $^{26}$Al 
in the solar system.

\section{Observations} \label{sec_obs}

\begin{figure}
\begin{center}
\includegraphics[width=\hsize]{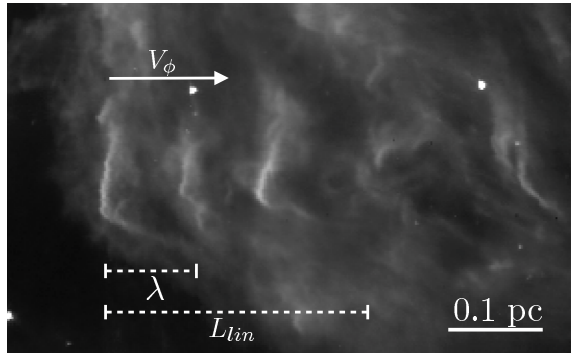}
\caption{ \emph{Spitzer} Infrared Array Camera \citep{faz04} 8 $\mu$m image of the \emph{Ripples}
in Orion which have been attributed to a KHI by BMC. $\lambda$ denotes the spatial wavelength of the structure,
$L_{lin}$ is the distance over which the instability appears to be completely linear. The orientation of $V_{\phi}$, the 
phase velocity of the KH wave at the surface of the cloud, is depicted qualitatively. \label{fig_obs}}
\end{center}
\end{figure}

Fig.~\ref{fig_obs} shows the \emph{Spitzer} Space Telescope \citep{wer04} 
mid-infrared (mid-IR) image of the \emph{Ripples} in Orion, which have been attributed 
to a KHI by BMC. KHIs occur at the interface between two fluids flowing relative
to each other. In the case of Orion, the two fluids are the \hii gas and the neutral 
gas of the molecular cloud, and the velocity sheer results from the \emph{champagne flow} 
created by the \hii region bursting through the parental molecular cloud. 
When growing, the KHI gives to the interface a wavelike structure which can be seen
in Fig. \ref{fig_obs}. An important parameter of the KHI is its spatial wavelength
$\lambda$ which is connected to the physical conditions in which the instability 
occurs. Here, $\lambda$ can be measured directly from the image and  
is found to be $\lambda=0.11$ pc for a distance to the Orion nebula of 414 pc. 
This corresponds to a spatial wavenumber $k=2\pi/\lambda=2\times10^{-15}$m$^{-1}$.

\section{Linear analysis of the Kelvin Helmholtz instability in Orion} \label{sec_mod}

\subsection{Objectives and method}

\begin{figure}
\begin{center}
\includegraphics[width=\hsize]{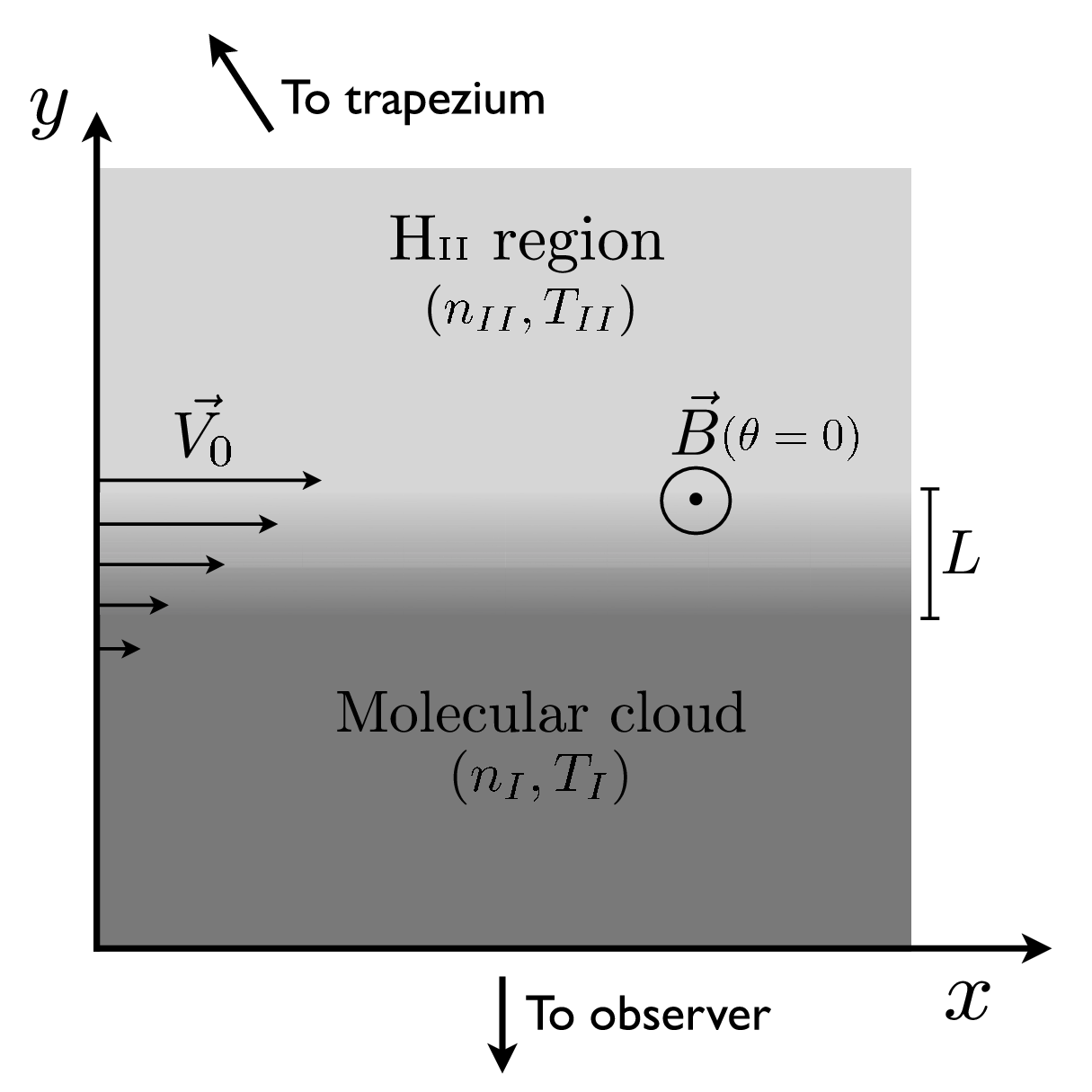}
\caption{Schematic representation of the configuration adopted for the initial conditions of the interface
between the \hii region and the molecular cloud. \label{fig_ske}}
\end{center}
\end{figure}

Our goal here is to study from the theoretical point of view, and using realistic physical 
conditions, the stability of an \hii / molecular cloud interface against the KHI. In particular
we want to determine wether magnetic fields can play a stabilizing role. It is also of 
great interest to derive some of the key parameters, for instance the growth rate $\gamma$ 
and tte range of acceptable wavelengths. In order to do 
this, we perform a linear stability study which includes magnetic field, compressibility and an analytical velocity
profile across the sheered layer (see below). This differs greatly from the preliminary 
study of BMC which relied on an ideal case  \citep{cha61} of incompressible 
fluids with a discontinuous velocity profile and no magnetic field. 
The present study is  performed in a two-dimensional slab 
geometry, described in Fig.~\ref{fig_ske} for the initial conditions. These initial conditions 
are maintained by the magnetohydrodynamical equilibrium. The density gradient and velocity
gradients are along the $y$ axis. The velocity and density profiles across the interface are 
of hyperbolic-tangent form \citep{miu82}. The velocity is oriented along the $x$ axis. The magnetic
field direction is inside the plane defined by $y$ and $z$ and its orientation is 
defined by the angle $\theta$ between $\vec{B}$ and $z$.  For this first analysis, we have not
considered the azymuthal dependance for the orientation of $\vec{B}$, because this would imply
heavy complications in the solving of the MHD equations. For the adopted configuration, the MHD equations 
are linearized and a perturbed quantity $f$ can be expanded as a plane wave in the form of 
$f(x,y) = \hat f(y)\exp\{i(k_x x-\omega t)\}$. The linearized equations can then be solved, with boundary 
conditions in the $y$ direction and a given wave number $k_x$ for the corresponding eigen value 
(angular frequency and growth rate) as an eigen value problem. This is described in mathematical terms 
in the following section.

\section{Linear model}
\subsection{Basic equations}
The basic MHD equations are
\begin{eqnarray}
\frac{\partial \rho}{\partial t} = -\nabla\cdot(\rho{\bf V}),
\label{eqn:continue}
\end{eqnarray}
\begin{eqnarray}
\frac{\partial {\bf V}}{\partial t} = -({\bf V}\cdot\nabla){\bf V} - \frac{1}{\rho}\nabla(P+\frac{B^2}{8\pi}) + \frac{1}{4\pi}({\bf B}\cdot \nabla){\bf B},
\label{eqn:moment}
\end{eqnarray}
\begin{eqnarray}
\frac{\partial P}{\partial t} = -({\bf V}\cdot \nabla)P - \Gamma P(\nabla \cdot {\bf V}),
\label{eqn:state}
\end{eqnarray}
\begin{eqnarray}
\frac{\partial {\bf B}}{\partial t} = -c \nabla \times {\bf E},
\label{eqn:faraday}
\end{eqnarray}
with the frozen-in condition
\begin{eqnarray}
{\bf E} = -\frac{\bf V}{c}\times{\bf B}.
\end{eqnarray}
where $\Gamma=5/3$ is a polytropic constant. The mass density $\rho$ and the magnetic field {\bf B} are normalized by characteristic values of $\rho_0$ and $B_0$, the velocity {\bf V} by the velocity jump across the boundary, $V_0$, the pressure P by $B_0^2/8\pi$, the spatial scale by the initial shear width $L$, and the time by $L/V_0$.

\subsection{Linearization and solution}
We consider perturbed quantities from an equilibrium state as, 
\begin{eqnarray*}
{\bf V} &=& {\bf V_0}+\delta {\bf v},\\
P &=& P_0+\delta p,\\
{\bf B} &=& {\bf B_0}+{\bf b}.
\end{eqnarray*}
The perturbed quantities are expressed as a plane wave in the form, $\delta A=\tilde A(y)\exp\left[{i(k_xx-\omega t)}\right]$, where $A$, $k_x$ and $\omega=\omega_r+i\gamma$ denote a physical parameter, the wave number in the x direction, and the angular frequency, respectively. 

Linearizing the above MHD equations (\ref{eqn:continue}-\ref{eqn:faraday}), one obtains
\begin{eqnarray}
\omega\delta v_x &=& k_xV_x\delta v_x - i\frac{\delta_Vx}{\partial y}\delta v_y + \frac{k_x}{2n_0}\delta p+ k_x\frac{B_{0z}}{n_0}b_z+\frac{i}{n_0}\frac{\partial B_{0x}}{\partial y}b_y \nonumber, \\
\omega\delta v_y &=& k_xV_x\delta v_y - \frac{i}{2n_0}\frac{\partial}{\partial y}(\delta p +2{\bf B_0}\cdot{\bf b}) - k_x\frac{B_{0x}}{n_0}b_y  \nonumber, \\
\omega\delta v_z &=& k_xV_x\delta v_z - k_x\frac{B_{0x}}{n_0}b_z + \frac{i}{n_0}\frac{\partial B_{0z}}{\partial y}b_y \nonumber, \\
\omega\delta p &=& k_xV_x\delta p - i\delta v_y\frac{\partial P_0}{\partial y}+\Gamma P_0k_x\delta v_x - i\Gamma P_0\frac{\partial \delta v_y}{\delta y}\nonumber, \\
\omega bx &=& k_xV_xb_x -iB_{0x}\frac{\partial\delta v_y}{\partial y} - i\delta v_y\frac{\partial B_{0x}}{\partial y}+ib_y\frac{\partial V_x}{\partial y} \nonumber, \\
\omega by &=& k_xV_xb_y - k_xB_{0x}\delta v_y \nonumber, \\
\omega bz &=& k_xV_xb_z + k_xB_{0z}\delta v_x -iB_{0z}\frac{\partial\delta v_y}{\partial y} - i\delta v_y\frac{\partial B_{0z}}{\partial y}-k_xB_{0x}\delta v_z.
\end{eqnarray}
Discretizing the spatial derivative of the Fourier amplitude in the y direction with boundary conditions at $y=\pm y_b$ far away from the shear layer, the equations cans be cast into the form of a eigen value problem as
\begin{equation}
\omega
\left(
\begin{array}{c}
\delta v_x\\
\delta v_y\\
\delta v_z\\
\delta p\\
b_x\\
b_y\\
b_z
\end{array}
\right )
=
M
\left(
\begin{array}{c}
\delta v_x\\
\delta v_y\\
\delta v_z\\
\delta p\\
b_x\\
b_y\\
b_z
\end{array}
\right ),
\end{equation}
from which we can obtain the eigen values of $\omega$, whose imaginary part is the growth rate, 
and the corresponding eigen functions for any $k_x$. To solve this eigen value problem, we use 
used the QR algorithm \citep{fra61,fra62,kub63}.

\subsection{Adopted initial parameters}

Solving the above described problem requires to know the initial conditions of the set-up.
Most of them can be derived from observations, or estimated. The adopted physical conditions 
are summarized in Table~\ref{tab_con}. The density and temperature ($n_I, T_I$)  of the 
molecular cloud have been discussed in BMC. According to their results we adopt 
$n_I=10^4$~cm$^{-3}$ and $T_I=20$~K. For the \hii region, we consider a typical temperature 
$T_{II}=10^4$~K, and $n_{II}=20$~cm$^{-3}$. 
The velocity sheer is taken to be $V_0=10$~km.s$^{-1}$ following \cite{roy95}, also typical for 
such environments.  Magnetic field strength has been measured in the Orion nebula by 
\citet{bro05} and \citet{abe04} and found to vary between 5nT for the Trapezium
region and and 25 nT in the Veil region. The \emph{Ripples} are likely situated between the 
Trapezium and the Veil so we adopt a conservative value of $B=20$ nT. 
The value of $\theta$ cannot be determine independently so we have considered various
values between 0 and 90$^{\circ}$ (with symmetrical results in the -90 to 0$^{\circ}$ range).
Finally, BMC derived a gravitational field at the cloud surface $g=3.5 \times 10^{-11}$ m~s$^{-2}$.
We can compare the potential to kinetic energy using the Richardson number $R_i=g~L/(V_0)^2$ \citep{cha61}.
The value of $L$ can be estimated to be the thickness of the photodissociation region
(PDR) measured by BMC, which is essentially the region where the gas
is converted from fully neutral to fully ionized and where the temperature changes from
a few 10 K to a few 1000 K \citep{tie85}. $L$ is typically 0.01 pc (BMC), 
hence $R_i\sim10^{-3}$ which implies that gravity is ineffective and it is no further included in 
the calculations.

\begin{table*}
\caption{Physical parameters}
\label{tab_con}
\begin{center}
\begin{tabular}{lccr}
\hline \hline
Parameter & Symbol &  & Reference \\
\hline
\multicolumn{4}{c}{Observational}\\
\hline
Heliocentric distance 		& $d$			& 414 pc					& (1) \\
Neutral gas density 			&$n_I$ 			& $10^4$ cm$^{-3}$ 		& (2) \\
Ionized gas density 			&$n_{II}$			& 20 cm$^{-3}$ 			&  (2) \\
Neutral gas Temperature		& $T_I$			& 20 K					& (3) \\
Ionized gas Temperature 		&$T_{II}$			& $10^4$K				& (4) \\
Velocity sheer adopted here 	&$V_0$			& 10 km.s$^{-1}$			&  (5) \\
Gravitational field 			&$g$			& 3.5 $\times10^{-11}$ m.s$^{-2}$& (2)\\
Magnetic field strength 		&$B$			& 20 nT					&  (6)\\
Magnetic field orientations 	& $\theta$			& 0-90$^{\circ}$			& \\
Instability wavelength 		&$\lambda$		& 0.11 pc 					&  (2)\\
Width of the sheered layer 	&$L$			& 0.01 pc					& (2) \\
Linear regime length		&$L_{lin}$		& 0.3 pc					& Fig. 1\\
\hline
\multicolumn{4}{c}{Derived values (for maximal growth rate)}\\
\hline
Instability growth rate		&$\gamma$		& 2.3$\times10^{-13}$s$^{-1}$	&  \\
Instability phase velocity		& $V_{\phi}$ 		& 3.6 km.s$^{-1}$ & \\
Instability saturation timescale	&$t_{sat}$			& 2$\times10^{5}$ yrs	&  \\
Size of mixing layer after $t_{sat}$ &$L_{mix}$		&0.5 pc	&  \\
Distance travelled before saturation 	& $L_{sat}$ 		&0.74 pc \\
\hline
\multicolumn{4}{p{13cm}}{(1) From \citet{men07}, (2) From BMC, (3) Typical for molecular clouds \citep{tie05}, (4) Typical for \hii regions \citep{tie05}, 
(5) From \citet{roy95}, (6) Based on \cite{abe04}  }\\
\end{tabular}
\end{center}
\end{table*}

\section{Results} \label{sec_res}

\subsection{Linear phase of the KHI}

\begin{figure}
\begin{center}
\includegraphics[width=\hsize]{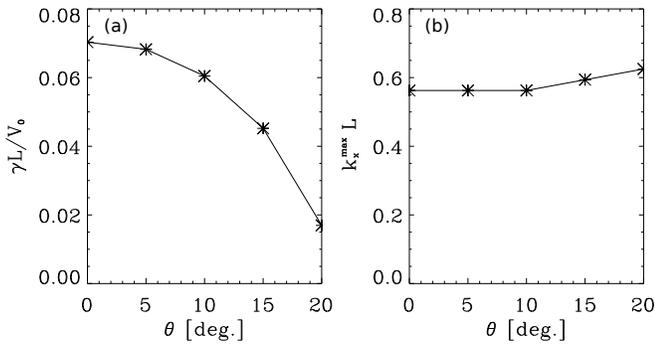}
\caption{\emph{Left:} Normalized growth rate $\gamma L/V_0$, of the most unstable mode, as a function
of the orientation of the magnetic field. \emph{Right:} Normalized spatial wavenumber $k_xL$ of the most 
unstable mode as a function of the orientation of the magnetic field.
 \label{fig_res}}
\end{center}
\end{figure}

\begin{figure}
\begin{center}
\includegraphics[width=\hsize]{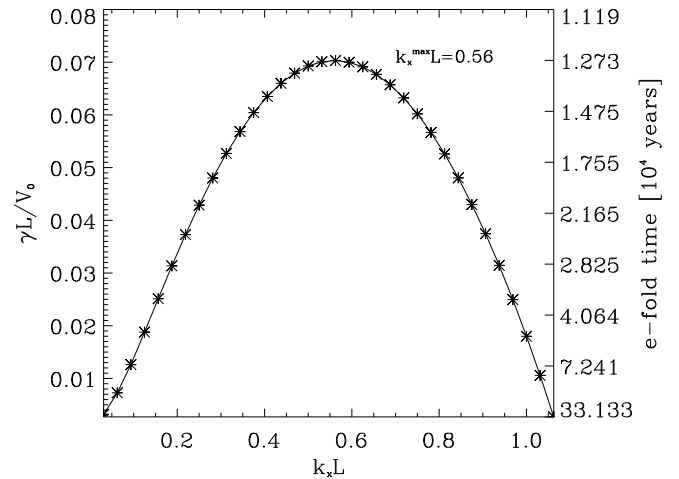}
\caption{ Normalized growth rate $\gamma L/V_0$ versus normalized spatial wavenumber $k_x L$ of the instability.
The right axis translates the value of the growth rate into an e-fold timescale for Orion. 
\label{fig_res2}}
\end{center}
\end{figure}

The results of the linear analysis concern the linear growth of the instability. They are presented 
in Fig.~\ref{fig_res} and Fig.~\ref{fig_res2}.
Fig.~\ref{fig_res} shows the influence of the value of $\theta$, the inclination of the magnetic field,
on the normalized growth rate $\gamma L / V_0$ and on the normalized wavenumber $k_xL$. 
The growth rate decreases with increasing angle but  for $-25^{\circ}\lesssim~\theta~\lesssim 25^{\circ}$ 
the growth rate is non-zero, implying that the interface is KH unstable. In this range of values for $\theta$, 
the wavenumber $k_x$ is almost constant, showing that it does not depend on magnetic field
orientation. Fig.~\ref{fig_res2} shows the evolution of the normalized growth rate $\gamma L / V_0$  
as a function of the normalized wavenumber $k_xL$, for $\theta=0$.
The most unstable mode corresponds to $k_xL=0.56$, and typically, for 
$0.1<k_xL<1$ the growth rate is high and corresponds to an e-fold timescale
of less than 10$^5$ years, short compared to the lifetime of an OB association ($\sim$10 Myrs). 
Therefore, it is expected that instabilities with $0.1<k_xL<1$ appear in star-forming
regions. Using $L=0.01$~pc, imposes that the wavelength of the instability  
$\lambda_{KH}$ will range between 0.06 and 0.6 pc. This number also places
limits on the detectability of KHIs: at a distance of 1 kpc this is an
angular size of 4-40'', and at 10 kpc this is 0.4-4''. Hence, KHI structures are expected 
to be of small angular size and can only be observed with high angular resolution
telescope and/or in nearby regions of massive star formation like Orion.

\subsection{Saturation of the KHI}

 \citet{mat10} studied in details the 2 dimensional 
 evolution of the KHI for conditions similar to those presented here and found that 
 the growth of the instability leads to saturation. This results 
 in the formation of a turbulent layer where the two fluids are mixed, over a time-scale 
 of the order of $t_{sat}\sim200L/V_0$. Hence, using the observed value for $L$ 
 (Table~\ref{tab_con}) this results in a saturation timescale $t_{sat}\sim 2\times10^5$ 
 years. It is important to realize that even if $t_{sat}$ is short there is always a 
 part of the instability that remains linear. According to the results of our linear analysis
 we find that the KH mode travels along the boundary 
 layer with speed of  $V_{\phi}=0.36\times V_0$, that is $\sim$ 3.6 km/s. Hence, we 
 can define $L_{sat}$, the spatial scale before the instability has saturated by 
 $L_{sat}=V_{\phi}~t_{sat}$ and find $L_{sat}\sim0.74$~pc.  
These theoretical results have several implications. First, the timescale for saturation is
short compared to the lifetime of an OB association, so KHIs will saturate and will 
be a source of turbulence. Secondly, the scale size over which it is possible 
to see the linear regime is \emph{at maximum} 0.74 pc.


\section{The \emph{Ripples} as a KHI}

Based on the theoretical results described above, we investigate in more details
if the \emph{Ripples} can be interpreted as an occurrence of the linear phase
of the KHI. The observed value of $\lambda$ for the \emph{Ripples} is 0.1 pc, 
which fall in the range defined from the theoretical investigation ($0.06<\lambda_{KH}<0.6$~pc).
The \emph{Ripples} seem to preserve a very periodic structure, suggesting linear regime, 
over a distance $L_{lin}=3\lambda=0.3$~pc (Fig. 1). The following 2 billows, instead, start 
showing some chaotic structure suggesting the beginning of the 
 saturation of the instability. After about 5$\lambda$, the periodic structure has disappeared. 
 Hence, the travelled distance in linear regime $L_{lin}$ for the \emph{Ripples} is, 
 consistently with the KHI model, smaller than the maximal theoretical value of $L_{sat}=0.74$~pc discussed in Sect.~\ref{sec_res}.
Therefore, we argue that the observed evolution of the \emph{Ripples} structure results from
 the motion of the KH wave towards saturation, from left to right in Fig. 1.
Our last remark concerns the importance of the magnetic field. As mentioned above
the \emph{Ripples} can only result from a KHI if they are in a region of small
$\theta$. Orientation of the magnetic field lines in Orion have been measured 
by e.g. \citet{hou04, poi10} using polarimetry. Unfortunately, this does not cover 
the \emph{Ripples} region and, in addition, performing such measurements at the 
arc-second  scale remains challenging. We can only stress the importance 
--often neglected-- that magnetic fields have in shaping the interstellar medium, 
in this case because they can stabilize interfaces between \hii regions and molecular 
clouds.

Altogether, we conclude that the study presented in this letter brings additional evidence 
that the \emph{Ripples} result from a KHI. This raises one question however, which is 
``why do we only see one occurrence of the KHI in Orion?'' This is
perhaps because only in this region are the conditions (e.g. magnetic field, velocity flow)
favorable at the moment we observe Orion. Over the lifetime of the region however, this
may occur a high number of times. Ripples have indeed been observed recently in an other
star forming region (Cygnus OB2) by \citet{sah12}, suggesting that the structure in Orion 
is not unique. It is also possible that more of these structures exist in Orion, but given 
their small angular size or unfavorable orientation on the plane of the sky they remain 
undetectable (see Sect. \ref{sec_res}).

\section{The KHI and chemical mixing in star-forming regions}

\subsection{Saturation of the instability: turbulence and chemical mixing}

We have discussed in Sect.~\ref{sec_res} the saturation of the KHI towards a turbulent
regime. This may have an important role in chemical mixing in star forming regions.
\citet{roy95} were the first to recognize the importance of the KHI in chemical mixing
of the interstellar medium.They studied the influence of KHIs by defining the e-fold timescale
assuming a spatial wavelength of 100 pc (the size of a large \hii region), and from this
derived a timescale of $1.5\times10^6$ years, which is smaller than the lifetime
of an OB association. However, their model was based on an ideal hydrodynamical 
 case with no magnetic field, for which the wavelength of maximal growth rate cannot 
 be determined. This is partly due to the fact that no observational evidence of
 KHIs existed at the time.  The results presented in this Letter using a more detailed model
 and guided by direct observations, are clearly incompatible with a maximal 
 growth rate corresponding to $\lambda=100$~pc. In addition, \citet{roy95} used the e-fold 
 timescale as a measure of mixing timescale, which is not appropriate. Instead,
 we can use the results found here and those of \cite{mat10} to obtain some general
 appreciation of the efficiency of the KHI mixing in star-forming regions. First, as mentioned
above, full mixing of the fluids occurs after $t_{sat}$ which we have found
 to be of the order of a few $10^5$ years.  Again, this is short compared to the lifetime of an 
 OB association so the process will be efficient.  After this time, the size of the mixed layer is 
 $L_{mix}=50\times L$ (\citealt{mat10} Fig. 9), that is 0.5 pc. All in all, in agreement
 with \citet{roy95} (although making different hypotheses), we conclude 
 that the KHI can be an efficient mechanism to mix chemical elements in 
 the interstellar medium.

 \subsection{Further implications: $^{26}$Al in the solar system}
 
 It is believed that low-mass star formation is triggered in the over-dense shell
 of molecular cloud that lie around  \hii regions (see e.g. \citealt{deh10}). In a recent paper, 
  \citet{gou12} argue that the Solar system may have formed in such an environment,
  based on the abundances of short lived radionucleides found in meteorites.
  In particular, they propose that $^{26}$Al was brought to the forming solar 
  system by the winds of a massive star (see also \citealt{mon07}) rather than by Supernovae. This requires 
  that a nearby massive star ($M_{\star}>32 M_{\sun}$) injected $^{26}$Al during
  a few million years, and that this element was then well mixed with the \hii 
  gas, and eventually that the \hii gas was mixed with the surrounding molecular shell 
  efficiently. In Orion at least, the gas from the wind seems to be well mixed
  with the \hii region as shown by \citet{gue08}. The efficiency of \hii mixing 
  with the molecular shell  was not evaluated by \citet{gou12}. However, they
  derive the time $t_{\star}$, during which the molecular shell has to be enriched
  before the solar system starts to form. This value ranges between 0.65 and 6.2 
  Myr. The mixing timescale we have derived here ($t_{sat}=2\times10^5$ yrs)
 is smaller than $t_{\star}$ so that indeed mixing by the KHI is efficient 
 to enrich the molecular shell with $^{26}$Al. Hence, the KHI (and possibly other instabilities) 
 could have played an important role in the transport of $^{26}$Al to the forming Solar system.
 
 \section{Conclusion}
 
We have shown that the KHI develops rapidly at the \hii molecular cloud interface
in conditions like Orion (which are representative of many massive star forming regions).
After travelling at the surface of the cloud during a time of a few 10$^5$ years and over a maximum 
distance of $\sim$ 0.74 pc, the instability reaches saturation. Hence, as suspected, the KHI is probably 
a significant mechanism to generate small scale ($<1$~pc) turbulence in molecular clouds near massive stars. 
In addition, since the \hii region is contaminated by the chemical elements injected by 
massive stars winds, the KHI may be a relevant process to bring these 
elements inside the molecular cloud, in regions where planetary systems around young 
stars are formed. This could have been at play to transport $^{26}$Al to the nascent
solar system. Periodic structures corresponding to the linear phase of the
KHI (like the \emph{Ripples}) should be relatively widespread in star-forming regions, for instance
on the surface of molecular globules as reported recently \citep{sah12}. However,  these structures 
are expected to be small (few arc-seconds) and hence require high angular resolution 
observations to be identified.

\acknowledgments 

This work was partly supported by Chiba University. We acknowledge 
Thierry Montmerle, Pierrick Martin, Tomoyuki Hanawa, and Ryoji Matsumoto 
for fruitful discussion. O. B. is funded by a CNES fellowship.

\bibliographystyle{apj}

\end{document}